\documentstyle[11pt,amssymb,amsfonts,amsthm,amsmath,amssymb,amscd,amstext,epsfig]{article}

\def\ben{\begin{equation}}
\def\een{\end{equation}}
\def\half{{\textstyle{1\over2}}}
\def\qtr{{\textstyle{1\over4}}}
\let\a=\alpha \let\b=\beta

\let\w=\omega

\let\pa=\partial
\def\be{\begin{equation}}
\def\ee{\end{equation}}
\def\ba{\begin{array}}
\def\ea{\end{array}}

\def\dalemb#1#2{{\vbox{\hrule height .#2pt
        \hbox{\vrule width.#2pt height#1pt \kern#1pt
                \vrule width.#2pt}
        \hrule height.#2pt}}}

\newcommand{\bea}{\begin{eqnarray}}
\newcommand{\eea}{\end{eqnarray}}

\def\R{{{\Bbb R}}}

\def\Z{{{\Bbb Z}}}

\def\ocal{{\mathcal{O}}}

\def\xp{{x^+}}
\def\xm{{x^-}}
\def\xb{{\boldsymbol{x}}}
\def\kb{{\boldsymbol{k}}}

\begin{document}

\begin{flushright}
NSF-KITP-08-124 \\
arXiv:0810.0298 [hep-th]
\end{flushright}

\begin{center}

\vspace{1cm} { \LARGE {\bf Families of IIB duals for nonrelativistic CFTs}}

\vspace{1cm}

Sean A. Hartnoll$^\flat$ and Kentaroh Yoshida$^\sharp$

\vspace{0.8cm}

{\it $^\flat$ Jefferson Physical Laboratory, Harvard University\\
     Cambridge, MA 02138, USA }

\vspace{0.7cm}

{\it $^\sharp$ Kavli Institute for Theoretical Physics, University of California \\
 Santa Barbara, CA 93106, USA } \\

\vspace{0.6cm}

 {\tt hartnoll@physics.harvard.edu, kyoshida@kitp.ucsb.edu} \\

\vspace{2cm}

\end{center}

\begin{abstract}

We show that the recent string theory embedding of a spacetime with
nonrelativistic Schr\"odinger symmetry can be generalised to a twenty
one dimensional family of solutions with that symmetry. Our solutions include
IIB backgrounds with no three form flux turned on, and arise as near horizon
limits of branewave spacetimes. We show that there
is a hypersurface in the space of these theories where an instability
appears in the gravitational description, indicating a phase transition
in the nonrelativistic field theory dual.
We also present simple embeddings of duals for nonrelativistic
critical points where the dynamical
critical exponent can take many values $z \neq 2$.

\end{abstract}

\pagebreak
\setcounter{page}{1}

\section{Background}

In recent studies of the (strongly coupled) quark-gluon plasma,
just above the QCD deconfinement temperature, the AdS/CFT
correspondence \cite{AdS/CFT} is finally starting to deliver
qualitative and even quantitative results of experimental
relevance that are difficult to obtain by more traditional
theoretical methods (for overviews see for instance
\cite{Mateos:2007ay, Son:2007zz, Gubser:2007zz}).

The best understood versions of the AdS/CFT correspondence
describe a duality between a conformally invariant gauge theory
and string theory in an asymptotically Anti-de Sitter space
background. Taking a large $N$ and (if necessary) strong coupling
limit in the gauge theory corresponds to taking the Anti-de Sitter
space to be weakly curved, and hence string theory may be
described by a ten or eleven dimensional supergravity theory.

QCD is a gauge theory, and so it is natural to hope that certain
quantities will be in the same `universality class' as the
supersymmetric gauge theories that can be studied using AdS/CFT.
This hope seems to be bourn out for the high temperature
deconfined phase of QCD \cite{Mateos:2007ay, Son:2007zz,
Gubser:2007zz}. On the other hand, there are many quantities of
interest that are sensitive to the fact that QCD is not a
conformal theory, but rather asymptotically free and confining.
Extensions of the AdS/CFT correspondence to non-conformal theories
are well studied. However, they generically suffer from the
feature that parametrically decoupling the confining physics from
other scales in the string background (such as the Kaluza-Klein
scale) requires going beyond the supergravity approximation. This
problem is reviewed for instance in \cite{Aharony:2002up}.

One is then led to wonder whether there are exactly conformally
(or at least, scale) invariant systems in nature that would be
kinematically closer to the simplest versions of AdS/CFT.
Conformal invariance has in fact been observed to arise in an
increasing number of condensed matter systems - in the vicinity of
a quantum critical point \cite{sachdev, sachdev2, heavy}.

The microscopic description of these condensed matter systems is
certainly not relativistic. It is remarkable therefore that
certain quantum critical points exhibit an emergence of the full
relativistic conformal symmetry. This occurs for instance in
certain superfluid-insulator transitions and at certain
`deconfined' quantum critical points (see for instance
\cite{Herzog:2007ij, Hartnoll:2007ih} for references). If a nearby relativistic
quantum critical point is believed to play a role in some
dynamical process, then one can hope that standard AdS/CFT
techniques can be applied. This was the methodological assumption
in \cite{Hartnoll:2007ih, Hartnoll:2007ip, Hartnoll:2008hs}, which
studied certain transport processes in unconventional
superconductors.

It is also common for quantum critical points to display a
non-relativistic scale invariance. These theories have a scale
invariance that treats space and time anisotropically, as captured
by the dynamical critical exponent $z$. Thus for instance, as the
critical point is approached, the mass gap $\Delta$, which
determines relaxation timescales, and the coherence length $\xi$
are related by
\be
\Delta \sim \frac{1}{\xi^z} \,.
\ee
The case $z=1$ corresponds to a relativistic theory. For general
$z$ the symmetry of the theory is simply the Galilean group
enhanced by a dilatation operation. The case $z=2$ is also singled
out, as the algebra may be enhanced in this case by an additional
`special conformal' generator. This is called the Schr\"odinger
algebra and enforces strong constraints on the dynamics. We will
not review this algebra in detail, for recent discussions see for
instance \cite{Nishida:2007pj, Son:2008ye,
Balasubramanian:2008dm}. Early discussion of the symmetry of the
Schr\"odinger equation are \cite{Hagen:1972pd, Niederer:1972zz}
and an early geometric discussion of these symmetries appears in
\cite{Duval:1990hj}. Examples of experimental systems
with $z=2$ include fermions at unitarity, see \cite{Son:2008ye,
Balasubramanian:2008dm} for references.

Recent work has suggested that the AdS/CFT correspondence can be
adapted to study strongly coupled non-relativistic quantum
critical points. We will concentrate in this paper
on the most constrained case of $z=2$, and on field theory in
$2+1$ dimensions. We
shall also present some interesting extensions of our results to
general $z$. The starting point is to write down a metric upon
which the Schr\"odinger symmetry is geometrically realised. The
following five dimensional metric does the job
\cite{Son:2008ye, Balasubramanian:2008dm}
\be\label{eq:background1}
ds^2_5 = R^2 \left( r^2 \left( - 2 d\xp d\xm + d\xb^2 \right) +
\frac{dr^2}{r^2} - \sigma^2 r^4 (d\xp)^2 \right) \,.
\ee
Here $\sigma$ is a parameter measuring deformation of the metric
away from Anti-de Sitter space. Furthermore, one should
periodically identify
\be\label{eq:periodic}
\xm \sim \xm + 2 \pi r^- \,.
\ee
This allows $P_- \equiv i \pa_\xm$ to be interpreted as particle
number
\be
P_- = \frac{N}{r^-} \,,
\ee
with $N$ an integer. If we then consider $\xp$ to be time, so that
$P_+ \equiv i \pa_\xp$ is energy, then positivity of energy for
classical fields in the spacetime (\ref{eq:background1}) requires
$N \geq 0$.

The works \cite{Son:2008ye, Balasubramanian:2008dm} showed that by
generalising standard AdS/CFT arguments to fields propagating in
the background metric (\ref{eq:background1}) one obtains sensible
results for correlators of operators in a Schr\"odinger invariant
theory. In order to exploit the full power of the AdS/CFT
correspondence, and obtain, for instance, the finite temperature
and number density backgrounds, it is necessary to embed the
metric (\ref{eq:background1}) into a consistent theory of gravity,
such as string or M theory. This was achieved in the works
\cite{Herzog:2008wg, Maldacena:2008wh, Adams:2008wt} who obtained
the following solution of IIB string theory on $AdS_5 \times X_5$,
with $X_5$ a Sasaki-Einstein space
\bea\label{eq:background2}
ds^2_{IIB} & = & R^2 \left( r^2 \left( - 2 d\xp d\xm + d\xb^2
\right) +
\frac{dr^2}{r^2} - \sigma^2 r^4 (d\xp)^2 + ds^2_{K-E} + \eta^2 \right) \,, \\
B_2 & = & \sigma e^{\Phi/2} R^2 r^2 d\xp \wedge \eta \,, \\
F_5 & = & 4 R^4 (1 + \star) \text{vol}(X_5) \,,
\eea
where $\Phi$ is the dilaton, which is constant. In this expression
the Sasaki-Einstein space has been written as a fibration over a
local K\"ahler-Einstein manifold
\be
ds^2_{X_5} = ds^2_{K-E} + \eta^2 \,,
\ee
such that $d \eta = 2 \omega$, the K\"ahler form on the
K\"ahler-Einstein base. In the maximally symmetric case $X_5 =
S^5$, the K\"ahler-Einstein manifold is ${\mathbb{CP}}^2$.
Closely related solutions had been found previously in
\cite{Alishahiha:2003ru, Gimon:2003xk}.

Before moving on, two comments are in order. Firstly, it was
emphasised in \cite{Maldacena:2008wh} that the circle that we have
periodically identified in (\ref{eq:periodic}) is null and
therefore has zero proper length. This means that strings winding
the circle are light and strictly speaking one cannot trust the
supergravity regime. It was also noted how this situation can be
ameliorated by considering finite density, as is indeed
appropriate for comparison to experimental
setups. Secondly, an alternative proposal for realising a gravity
dual of non-relativistic theories was presented in
\cite{Goldberger:2008vg, Barbon:2008bg}. These papers use
pure AdS, with the relativistic symmetry being broken by boundary
conditions. This approach has been critiqued, in our view
correctly, in \cite{Maldacena:2008wh, Herzog:2008wg}.

In this work we shall present some new solutions to IIB
supergravity with a Schr\"odinger symmetry. There will be a large
family of solutions (21 dimensional in the case of $AdS_5 \times
S^5$) that interpolate from the solution (\ref{eq:background2}) to
a solution which only involves the metric. As we move along the
space of solutions, an instability develops on a critical
hypersurface in parameter space. This instability is presumably
dual to a (quantum) phase transition in the coupling space of the
quantum field theories.

The single step in our construction of these solutions is to start with
$AdS_5 \times X_5$ and then to introduce a harmonic function
on $X_5$ into the $AdS_5$ part of the metric. An intriguing
feature of this construction is that the resulting exponent $z$ is directly
given by an eigenvalue of the Laplacian on $X_5$.
In general $z \neq 2$. We go on
to show how such spaces arise as the near horizon limit of a branewave
traveling along $N$ D3 branes. We also show that despite
the mixing of $X_5$ and $AdS_5$ coordinates, a scalar field
propagating in the ten dimensional background is
equivalent (in the $z=2$ case) to a scalar field
in the five dimensional space (\ref{eq:background1})
but with a renormalised mass.

Towards the end of this paper we construct two types of
explicit renormalisation group flow solutions that are flows away
from an ultraviolet (UV) Schr\"odinger invariance.
The first is a novel solution that appears
to describe our field theories with a finite energy density.
The second are nonrelativistic deformations of the known
Coulomb branch solutions of ${\mathcal{N}}=4$ super
Yang-Mills theory. We conclude with comments on the dual
field theory interpretation of our solutions, and a rather long
to do list.

\section{A family of IIB Schr\"odinger solutions}

\subsection{Solutions with no fluxes}

Consider the following IIB ansatz, which has the full
Schr\"odinger symmetry
\bea\label{eq:background3}
ds^2_{IIB} & = & R^2 \left( r^2 \left( - 2 d\xp d\xm + d\xb^2
\right) +
\frac{dr^2}{r^2} - f(X_5) r^4 (d\xp)^2 + ds^2_{X_5} \right) \,, \\
F_5 & = & 4 R^4 (1 + \star) \text{vol}(X_5) \,.
\eea
Here $X_5$ can be any five dimensional Einstein manifold, i.e.
$R^{X_5}_{mn} = 4 \, g_{mn}^{X_5}$. What we have done is to insert
a function $f(X_5)$, depending on the coordinates on $X_5$, into
the noncompact part of the metric. Thus the spacetime describes a
wave propagating in $AdS_5$ with a profile along the internal
$X_5$ directions. A similar ansatz was considered in
\cite{Kumar:2004jv}. In fact, section 2.2 of their paper describes
a specific example of the solutions we shall discuss in the
following subsection.

The wave in $AdS_5$ in (\ref{eq:background3}) only alters the
$R_{++}$ component of the Ricci tensor relative to the pure
$AdS_5$ background. It is easy to check that our ansatz solves the
Einstein frame IIB equations of motion
\be\label{eq:IIBeqns}
R_{MN} = \frac{1}{96} F_{M ABCD} F_N{}^{ABCD} \,,
\ee
provided that
\be\label{eq:harmonic}
- \nabla^2_{X_5} f = 12 f \,.
\ee
If $f$ diverges at any point this will introduce singularities
into the spacetime. These are not curvature singularities, but
rather infinite tidal forces that appear for geodesic observers.
For a recent discussion see for instance \cite{Marolf:2002bx}.
This is sufficient for time evolution in the spacetime to be
ill-defined, so we should require that $f$ is regular. Therefore
(\ref{eq:harmonic}) implies that $f$ should be a scalar harmonic
on $X_5$ with a specific eigenvalue.

When we consider general $z$ below, we will find that the scalar
harmonic eigenvalue depends on $z$. For the moment we can ask
which Einstein manifolds have a scalar harmonic with eigenvalue
12, as required by (\ref{eq:harmonic}). For $S^5$ we have the
spectrum
\be\label{eq:eigenvalues}
- \nabla^2_{S^5} = \ell (\ell + 4) \,.
\ee
So the $\ell=2$ harmonics give solutions. The degeneracy of $\ell
= 2$ on $S^5$ is 20, so $AdS_5 \times S^5$ has a twenty
dimensional family of solutions of the form
(\ref{eq:background3}). The degeneracy can be obtained as the
number of tracefree quadratic forms in six variables.

In fact, any Sasaki-Einstein manifold has eigenvalues of the form
(\ref{eq:eigenvalues}), see for instance \cite{Gauntlett:2006vf}.
However, in general $\ell$ does not run over all the positive
integers, so we are not guaranteed the existence of a mode with
eigenvalue 12. A familiar case where the spectrum is again known
exactly is the homogeneous Sasaki-Einstein manifold $T^{1,1}$. The
spectrum is (see e.g. \cite{Gubser:1998vd})
\be
- \nabla^2_{T^{1,1}} = 6 \left( \ell_1(\ell_1+1) +
\ell_2(\ell_2+1) - \frac{r^2}{8}
\right) \,.
\ee
Here $r$ is an integer, and $\ell_1, \ell_2$ can be integers or
half integers, constrained as specified in for instance
\cite{Gubser:1998vd, Gibbons:2004em}. There are precisely two combinations with
the required eigenvalue: $(r,\ell_1,\ell_2)=(0,1,0)$ and
$(0,0,1)$. Because $\ell_1$ and $\ell_2$ label standard spherical
harmonics on $S^2$, this means that we have a total of six modes
in this case, and hence a six parameter family of solutions.

It is interesting to note that if an Einstein manifold admits
eigenvalues with $-\nabla^2_{X_5}=12$, then the corresponding
$AdS_5 \times X_5$ compactification has a scalar mode in $AdS_5$
which saturates the Breitenlohner-Freedman bound for stability.
This is because certain metric and five form fluctuations about
$AdS_5 \times X_5$ have masses $m^2 R^2 = - \nabla^2_{X_5} + 16
\pm 8 \sqrt{-\nabla^2_{X_5}+4}$ (see e.g. \cite{DeWolfe:2001nz}).
Putting $-\nabla^2_{X_5}=12$ leads to the Breitenlohner-Freedman
bound for $AdS_5$: $m^2 R^2 = -4$.

On any $X_5$, a nontrivial spherical harmonic has regions in which
it is both positive and negative. This follows from orthogonality
of the harmonic with a constant function. Therefore our function
$f(X_5)$ appearing in the spacetime (\ref{eq:background3}) will be
negative in certain regions of the internal space. One might
therefore worry about stability
\cite{Marolf:2002bx, Brecher:2002bw} and geodesic completeness
\cite{Hubeny:2002zr} of the spacetime. We will return to these
points below. First we shall show how NSNS flux can be added to
our solution (\ref{eq:background3}) to allow an interpolation
between these solutions without flux and the known solutions
(\ref{eq:background2}).

\subsection{Solutions with fluxes}

We can now add a NSNS two form potential to our solution
(\ref{eq:background3}). This will allow the family of solutions to
connect onto the solutions (\ref{eq:background2}). The ansatz is
now
\bea\label{eq:background4}
ds^2_{IIB} & = & R^2 \left( r^2 \left( - 2 d\xp d\xm + d\xb^2
\right) +
\frac{dr^2}{r^2} - f(X_5) r^4 (d\xp)^2 + ds^2_{K-E} + \eta^2 \right) \,, \\
B_2 & = & \sigma e^{\Phi/2} R^2 r^2 d\xp \wedge \eta \,, \\
F_5 & = & 4 R^4 (1 + \star) \text{vol}(X_5) \,.
\eea
Here we again require $X_5$ to be Sasaki-Einstein and use the same
notations as above.

It is easy to check that the five form and three form field
equations are satisfied exactly as they were for the solution
(\ref{eq:background2}), the function $f(X_5)$ does not appear in
these equations. The Einstein equation becomes
\be\label{eq:einstein2}
R_{MN} = \frac{1}{96} F_{M ABCD} F_N{}^{ABCD} +
\frac{e^{-\Phi}}{4} \left(H_{MAB} H_N{}^{AB} - \frac{1}{12} g_{MN} H_{ABC} H^{ABC} \right)\,,
\ee
where $H_3=dB_2$ and $\Phi$ is the (constant) dilaton. The last
term in this equation vanishes because $H_3$ is null in
(\ref{eq:background4}). The Einstein equation is solved provided
that
\be
-\nabla^2_{X_5} f = 12 f - 12 \, \sigma^2 \,.
\ee
From this equation we obtain
\be
f = \sigma^2 + \tilde f \,,
\ee
where $\tilde f$ is, as previously in (\ref{eq:harmonic}), a
scalar harmonic on $X_5$ with eigenvalue $12$. We see that these
solutions are precisely a linear superposition of our solution
(\ref{eq:background3}) without fluxes, and the known solution
(\ref{eq:background2}). By tuning $\sigma$ and the magnitude of
$\tilde f$ we interpolate between these two limits. Thus we obtain
a 21 dimensional family of solutions from $AdS_5 \times S^5$ and a
7 dimensional family from $AdS_5 \times T^{1,1}$. We can also note
that by taking $\sigma$ to be sufficiently large relative to
$\tilde f$, the function $f$ appearing in the metric will become
positive everywhere. This fact will shortly have consequences in
our stability analysis.

\section{Scalar field fluctuations}
\label{sec:scalar}

One might have expected that the mixing of internal and noncompact
coordinates in our solutions (\ref{eq:background4}) would
complicate the study of fluctuations about the background. This is
not the case. We will now show that, for scalar fields at least,
the only effect of the mixing is to `renormalise' the effective
Kaluza-Klein mass of the modes.

The linearised equation of motion for a (massive) ten dimensional
scalar field is
\be\label{eq:laplace}
\nabla^2 \phi = m^2 \phi \,.
\ee
It is easily seen that we can solve equation (\ref{eq:laplace}) by
writing
\be
\phi = \Phi(r) Y_{\lambda}(X_5)\, e^{-i \w \xp + i \kb \cdot \xb - i M \xm} \,.
\ee
Here $\w$ is the frequency, $\kb$ the wavevector and $M$ the rest
mass. The eigenvalue $\lambda$ is the solution to a
Schr\"odinger-like equation on $X_5$
\be\label{eq:operator}
\left[ - \nabla^2_{X_5} + M^2 f \right] Y_\lambda = \lambda \, Y_\lambda \,.
\ee
The function $\Phi(r)$ solves
\be\label{eq:5d}
\frac{d^2\Phi}{dr^2} + \frac{5}{r} \frac{d \Phi}{dr} + \frac{2 M \w - \kb^2}{r^4} \Phi
= \frac{m^2 + \lambda}{r^2} \Phi \,.
\ee
Therefore we see that the only effect of the mixing of internal
and noncompact coordinates is to `renormalise' the effective mass
of the five dimensional field: $m^2 \to m^2 + \lambda$.
The wave equation separates even though the background is
not a direct product. If the
rest mass $M=0$, then there is no dependence on $f$ and
$Y_\lambda$ are simply the spherical harmonics on $X_5$. For
nonzero $M$, the functions $Y_\lambda$ solve a modified equation
on $X_5$.

The five dimensional equation (\ref{eq:5d}) is easily solved
\cite{Son:2008ye, Balasubramanian:2008dm} to give,
up to normalisation, the modified Bessel function
\be
\Phi = \frac{K_{\nu} (p/r)}{r^2} \,,
\ee
where
\be\label{eq:nu}
p = \sqrt{\kb^2 - 2 M \w} \,, \qquad \nu = \sqrt{m^2 + \lambda +
4} \,.
\ee
We have imposed regularity of the solution at the horizon $r=0$
(assuming that $p$ is real). The different modes of this equation
correspond to scalar operators $\ocal$ with scaling dimension
\be\label{eq:delta}
\Delta_\ocal = 2 + \nu \,.
\ee
If $0 < \nu < 1$ we can also take $\Delta_\ocal = 2 - \nu$
\cite{Son:2008ye}. Thus we see explicitly that the effect of our
function $f$ is to shift the scaling dimension of the operators
dual to bulk scalar fields.

\section{Stability and geodesic completeness: a phase transition}

We noted that in our solution without flux (\ref{eq:background3}),
the function $f$ appearing in the metric would necessarily be
negative over some region of $X_5$. This is potentially associated
with various spacetime pathologies, see e.g. \cite{Marolf:2002bx,
Brecher:2002bw}, two of which we shall now investigate. On the
other hand, in the previously known background
(\ref{eq:background2}), $f$ is positive everywhere. In our family
of solutions (\ref{eq:background4}) interpolating between these
limits, $f$ becomes negative on some critical hypersurface in the
space of solutions. A pathology arising on this surface would
indicate an interesting zero temperature phase transition in the
dual family of nonrelativistic field theories. We will see that
while the space remains geodesically complete across the
transition to nonpositive $f$, there is indeed an instability
once $f$ becomes sufficiently negative.

\subsection{Appearance of an unstable scalar fluctuation}

The first question we consider is stability against scalar field
perturbations, such as the dilaton. Stability in pp wave
backgrounds is complicated by various factors such as the lack of
global hyperbolicity \cite{Penrose:1965rx}. See however the
comments in, for instance,
\cite{Brecher:2002bw}. We shall take the following simple
approach that seems appropriate for the present context.

Firstly, we fix the sign of the rest mass $M \geq 0$. This is a
choice of direction of the Killing vector generating translations
in $\xm$. Next, perform a simple change of variables on the
effective five dimensional equation for a scalar field in
(\ref{eq:5d}) above. The objective is to put the equation in
Schr\"odinger form. Let
\be
u = \frac{1}{r} \,, \qquad \Phi = u^{3/2} \Psi \,.
\ee
Furthermore, we will put $\kb=0$, as zero momentum gives the
potentially most unstable mode. Then from (\ref{eq:5d}) we obtain
\be\label{eq:schro}
- \frac{d^2\Psi}{du^2} + \left( \nu^2 - \frac{1}{4} \right)
\frac{1}{u^2} \Psi = 2 M \w \Psi
\,.
\ee
Recall that $\nu$ is given by (\ref{eq:nu}) above. This is a very
well studied Schr\"odinger equation (with `Energy' $E = 2 M \w$).
In particular, it is easy to see that when $\nu^2 < 0$ there is a
continuum of bound states with arbitrarily low energies.
This means that there are normalisable solutions to
(\ref{eq:schro}) with arbitrarily negative $M \w$.

An instability is usually signalled in general relativity by an
exponentially growing mode, in which $\w$ is imaginary. This is
not what we have here. In the usual case of pure AdS space, $f=0$,
the solutions we have found for $\nu^2 < 0$ correspond to modes
with masses below the Breitenlohner-Freedman bound
\cite{Breitenlohner:1982jf}, that are growing exponentially in the
boundary Minkowski time $t = \xp + \xm$. In the present setup, we
have periodically identified $\xm$, and hence the rest mass $M$
must be real. This constraint, and self-adjointness of the
Schr\"odinger equation, requires $\w$ to be real also and thus
there is no exponential growth.

However, given that we have taken $M>0$, then $M\w$ negative
requires $\w$ negative. We will see shortly that large $M$ is
necessary to obtain $\nu^2 < 0$. Nonetheless, we can take $M$
large but finite. The fact that $M\w$ is unbounded below then
implies that we have normalisable solutions with $\w$ arbitrarily
negative. The frequency $\w$ is nothing but the energy of the dual
field theory, with respect to the nonrelativistic time $\xp$. Our
analysis will therefore imply that the Hamiltonian of the dual
nonrelativistic theory becomes unbounded below when $\nu^2 < 0$.
This is indeed an instability. We do not see exponential growth
because we have taken a free scalar field. Generic interactions
will send the system into energetic free-fall.

We can also note from (\ref{eq:delta}) above that $\nu^2 < 0$
corresponds to an imaginary scaling dimension for the dual
operator $\ocal$. This reinforces the idea that the system is
unstable in these cases.

The question is therefore: when do modes appear with $\nu^2 < 0$?
Recall that $\nu^2 = m^2 + \lambda + 4$. The standard $AdS_5 \times
X_5$ vacua of IIB string theory are stable provided that there are
no five dimensional modes with masses below the
Breitenlohner-Freedman bound: $m^2 + 4 \geq 0$. Let us assume we
are compactifying on such an $X_5$. This class of five manifolds
includes all Sasaki-Einstein manifolds. Our solutions will
therefore become unstable if $\lambda$ can become sufficiently
negative. Recall that the allowed values of $\lambda$ are
eigenvalues of the operator (\ref{eq:operator}) on $X_5$. It is
clear that if $f$ is positive in (\ref{eq:operator}) then
$\lambda$ must also be positive. Thus
\be\label{eq:stab}
f > 0 \quad \Rightarrow \quad \text{Stability.}
\ee
Of course we have only shown stability against the particular mode
(\ref{eq:laplace}).

We will now show that if we are sufficiently close in parameter
space to the solution (\ref{eq:background3}), with no three form
flux, then by taking $M$ sufficiently large an instability will
appear. We will not show this in generality, but rather for the
case of $S^5$. Let us scale $M$ to be parametrically large, which
we can do, as the theory contains excitations with arbitrarily
large particle number. It then follows from (\ref{eq:operator})
that a sufficient condition for an arbitrarily negative $\lambda$ is if we can
find a test function $Y$ on $S^5$ such that the potential energy
is negative. That is
\be
\exists Y \quad \text{s.t.}
\quad \int_{S^5} f Y^2 d\Omega \, < \, 0 \quad \Rightarrow \quad \text{Instability.}
\ee
The potential term will eventually dominate over the kinetic term
in (\ref{eq:operator}) if we take $M$ to be sufficiently large.

If we introduce cartesian coordinates $\{x^i\}$ on $\R^6$, then we
have from above that
\be\label{eq:func}
f = \sigma^2 + a_{ij} x^i x^j \,,
\ee
with $a_{ij}$ a symmetric tracefree real matrix. The function $f$
is of course restricted to the unit $S^5 \subset \R^6$. Let us now
take as a test function an $\ell = 1$ harmonic on the five sphere.
That is: $Y = b_k x^k$. This gives
\bea
\int_{S^5} f Y^2 d\Omega & = & \sigma^2 b_k b_l \langle x^k x^l
\rangle_{S^5} + a_{ij} b_k b_l \langle x^i x^j x^k x^l
\rangle_{S^5}\nonumber \\
 & = & \text{Vol}(S^5) \left(\frac{\sigma^2 b_i b_i}{6} + \frac{b_i a_{ij} b_j}{24} \right) \,.
\eea
Denote the eigenvalues of $a_{ij}$ by $a^{(i)}$. A tracefree
symmetric matrix necessarily has at least one negative eigenvalue.
Therefore
\be\label{eq:instab}
4\, \sigma^2 \, < \, |\min_i a^{(i)}| \quad \Rightarrow \quad
\text{Instability.}
\ee
This relationship shows that if $\sigma$ (i.e. the three form
flux) is sufficiently small, the theory becomes unstable, as
advertised. It may well be that the theory becomes unstable at a
larger value of $\sigma$ than that given in (\ref{eq:instab}). The
results (\ref{eq:stab}) and (\ref{eq:instab}) leave the stability
in the range $\sigma^2 \in \Big[\qtr |\min_i a^{(i)}|,|\min_i
a^{(i)}|\Big]$ undetermined.

To summarise the results from this section: Equations
(\ref{eq:stab}) and (\ref{eq:instab}) give conditions on the
function appearing in
(\ref{eq:func}) to give stable and unstable spacetimes, respectively.
We have not found the precise hypersurface in
$\{\sigma^2,a_{ij}\}$ space where an instability sets in,
but we have bounded it on both sides.

It is worth highlighting that this instability is catastrophic for
the spacetime. As soon as the the criterion $\sigma \leq
\sigma_\text{crit.}(a^{(i)})$ is satisfied then all modes in the theory
with sufficiently large particle number become unstable at once.
It would be extremely interesting
to understand the field theoretic dual description of this
process.

\subsection{Geodesic completeness}

Another potential pathology of the spacetime that can arise when
$f$ is not positive is geodesic incompleteness. In particular, we
have to check that no timelike or null geodesics can reach
infinity in a finite affinely parametrised worldline time
\cite{Hubeny:2002zr}. For simplicity let us consider a geodesic
that is at a constant location on $X_5$. This is possible at the
stationary points of $f(X_5)$. It is easy to check that relaxing
this restriction will not alter our conclusions -- the geodesic
would simply oscillate in the $X_5$ directions as it moves out
towards infinity.

It is straightforward to check that to leading order at large $r$,
(timelike or null) geodesics satisfy
\be\label{eq:geodesic}
\left( \frac{dr}{ds} \right)^2 \sim - (M^2 + f E^2) r^2 \,,
\ee
where $s$ is an affine parameter and $M^2$ and $E^2$ are positive
constants ($M$ can be zero). If $f$ is positive everywhere, then
this equation is inconsistent and hence geodesics cannot reach
infinity at all. If $f$ is negative then, by taking $E$
sufficiently large, the right hand side of (\ref{eq:geodesic}) can
become positive. The equation is easily solved to give
\be
r \sim e^{m s} \,,
\ee
for some positive constant $m$. Therefore we see that it takes an
infinite affine time for geodesics to reach infinity. The
spacetime is geodesically complete in all cases. This result is
closely related to the regularity of plane wave spacetimes
\cite{Horowitz:1989bv}. We shall see the connection to plane waves
explicitly in section \ref{sec:branewaves} below.

\section{Many solutions with $z \neq 2$}

Generalising our above construction to a large number of values of
$z$ is straightforward and leads to an interesting observation.
Consider the ansatz
\bea\label{eq:background5}
ds^2_{IIB} & = & R^2 \left( r^2 \left( - 2 d\xp d\xm + d\xb^2
\right) +
\frac{dr^2}{r^2} - f(X_5) r^{2z} (d\xp)^2 + ds^2_{K-E} + \eta^2 \right) \,, \\
F_5 & = & 4 R^4 (1 + \star) \text{vol}(X_5) \,.
\eea
Note that we have not included a $B_2$ field. The $B_2$ equation
is not easy to satisfy for general $z$, which is one of the reasons why the papers
\cite{Herzog:2008wg, Maldacena:2008wh, Adams:2008wt} concentrated
on the case $z=2$. With the above ansatz (\ref{eq:background5})
the Einstein equation is satisfied provided that
\be\label{eq:generalz}
- \nabla^2_{X^5} f = 4 (z^2-1) f \,.
\ee
Once again, we have regular solutions whenever the manifold $X_5$
admits a scalar harmonic with a specific eigenvalue.

For the $AdS_5 \times S^5$ case, equation (\ref{eq:generalz}),
together with (\ref{eq:eigenvalues}),
implies that we have solutions for all
\be\label{eq:zell}
z = \frac{\ell+2}{2} \,.
\ee
That is to say, $z$ can take an integer or half-integer value
greater than one. The dimensionality of the family of solutions
obtained is given by the number of tracefree rank $\ell$ symmetric
tensors on ${\mathbb{R}}^6$:
\be
\#(z) = \frac{1}{12} (1+\ell) (2 + \ell)^2 (3 + \ell) = \frac{z^2 (4 z^2 - 1)}{3} \,.
\ee
Clearly this becomes very large for large $z$.

For a general $X_5$, the allowed values of $z$ will be a sequence
of irrational numbers, directly related to the scalar eigenvalues
of $X_5$ through (\ref{eq:generalz}). This gives a rather
intriguing connection between spectral properties of Einstein
manifolds and dynamical critical exponents. A general statement
that can be made is the following. The Lichnerowicz theorem, see
e.g. \cite{Duff:1986hr}, applied to five dimensions, says that the
lowest nontrivial Laplacian eigenvalue of a five dimensional
Einstein manifold with $R_{mn}^{X_5} = 4 g_{mn}^{X_5}$ is bounded
below
\be
-\nabla^2_{X_5} \geq 5 \,,
\ee
with equality only for the case of $S^5$. Using our relation
(\ref{eq:generalz}) we obtain
\be\label{eq:zbound}
z \geq \frac{3}{2} \,.
\ee
Of course there is also the $z=1$ possibility of pure AdS which
corresponds to the trivial (zero) eigenvalue. Thus our
construction will only enable us to obtain theories with dynamical
critical exponents satisfying (\ref{eq:zbound}).

We should note the following two facts. Firstly, for $z > 2$ the
spacetime (\ref{eq:background5}) is not geodesically complete. It
is easy to see that at large $r$, and in regions of $X_5$ where
$f$ is negative, there are (null and timelike) geodesics that
behave like
\be
r \sim \frac{1}{(s-s_0)^{1/(z-2)}} \,.
\ee
Again, $s$ is an affine parameter and $s_0$ a constant. These
geodesics reach infinity at a finite affine parameter $s=s_0$. The
geodesic incompleteness may possibly be related to the lack of scaling
near infinity for $z>2$ that was noted in
\cite{Balasubramanian:2008dm}. It is not clear at this stage how
important this result is given that the large radius region is
problematic in any case \cite{Maldacena:2008wh}. This is because,
even with a putative solution at finite number density, at large
radius the periodic identification of the $x^-$ circle means that
wound strings become arbitrarily light and the supergravity
description breaks down. In our solution of course, without number
density, the circle is null everywhere.

Secondly, if $z < 2$ the spacetime is singular at the origin
$r=0$. Although the curvatures are finite, there are pp singularities
in regions where $f$ is negative.
For instance, a radially infalling null geodesic experiences
infinite tidal forces as $r \to 0$ in a parallelly propagating
orthonormal frame. These are mild null singularities that may
well disappear in a finite temperature solution.

We will not study these solutions further here. It would be
interesting to investigate their stability. Our experience in the $z=2$
case above and the absence of flux in these solutions might suggest that they
are unstable. However, unlike we found in section \ref{sec:scalar}
above, the ten dimensional wave equation in the background
(\ref{eq:background5}) does not separate nicely for $z \neq 2$,
which complicates the analysis.

\section{Undoing the near horizon limit: branewaves}
\label{sec:branewaves}

It is instructive to write down a IIB geometry that asymptotes to
a pp wave in flat space and which develops a Schr\"odinger
symmetry in a near horizon limit. The solution is
\bea\label{eq:branewave}
ds^2_{IIB} & = & \frac{1}{H(r)^{1/2}} \left( - 2 d\xp d\xm +
d\xb^2 - f(X_5) r^2 (d\xp)^2 \right) +
H(r)^{1/2} \left( dr^2  + r^2 ds^2_{X^5} \right) \,, \\
ds^2_{X^5} & = & ds^2_{K-E} + \eta^2 \,, \\
B_2 & = & \sigma e^{\Phi/2} r^2 d\xp \wedge \eta \,, \\
F_5 & = & \frac{-H'(r)}{H(r)^2} (1 + \star) dr \wedge dx^+ \wedge
dx^- \wedge dx^1 \wedge dx^2 \,.
\eea
Here $H(r)$ is the usual extremal D3 brane function
\be
H(r) = 1 + \frac{R^4}{r^4} \,.
\ee
It is straightforward to check that the equation for the three
form field strength is independent of the function $H$, and
therefore remains satisfied for this more general metric ansatz.
The Einstein equations (\ref{eq:einstein2}) are satisfied provided
that, once again,
\be
-\nabla^2_{X_5} f = 12 f - 12 \, \sigma^2 \,.
\ee
As above, this is solved by letting $f = \sigma^2 + \tilde f$,
with $\tilde f$ being a scalar harmonic on $X_5$ with eigenvalue
12. The near horizon limit is now $r \ll R$, which leads to the
solution we found in previous sections.

The solution (\ref{eq:branewave}) has an immediate interpretation
as a D3 brane in a pp wave background. In fact, in
\cite{Hubeny:2002nq} a class of solutions were found that includes
(\ref{eq:branewave}) in the case with no $B_2$. Similar solutions
were discussed in \cite{Brecher:2000pa}, although these had a wave
profile along the D3 brane directions. In the case we are
studying, the pp wave is travelling along the D3 brane, but has a
profile orthogonal to the D3 brane. The connection to D3 branes in
a pp wave background can be made especially explicit in the case
of $X_5 = S^5$, as the geometry can be rewritten as
\be
ds^2_{IIB} = \frac{1}{H(r)^{1/2}} \left( - 2 d\xp d\xm + d\xb^2 -
[\sigma^2 \delta_{i j} + a_{i j}] y^i y^j (d\xp)^2 \right) +
H(r)^{1/2} d{\boldsymbol{y}}^2
\,.
\ee
Here $\boldsymbol{y}$ are the six coordinates transverse to the D3
brane worldvolume, $r^2 = {\boldsymbol{y}}^2$, and $a_{ij}$ is a
tracefree matrix of real numbers. We used the fact that the
$\ell=2$ harmonics on $S^5$ are inherited from quadratic
polynomials in ${\mathbb{R}}^6$. This spacetime asymptotes to a
homogeneous ten dimensional plane wave. The near horizon
Schr\"odinger group can be thought of as a remnant of the large
symmetry of the plane wave. The tracefree part of the waveform
originates from a purely metric ten dimensional solution, whereas
the trace is supported by the null three form flux.

\subsection{Supersymmetries}

Given that the $D3$ brane and the homogeneous pp wave (without
flux, $B_2 = 0$) both preserve half of the supersymmetries of IIB
supergravity, it is natural to expect that this intersection of
the two solutions will preserve one quarter.

It is straightforward to check that our solution
(\ref{eq:background3}) with $B_2 = 0$ indeed preserves a quarter
of the supersymmetries. In the absence of NSNS three form flux,
all that is needed is that the gravitino variation vanish
\be\label{eq:killing}
\delta \psi_M = \left(D_M +
\frac{i}{192} \Gamma^{NPQRS} \Gamma_M F_{NPQRS} \right) \epsilon = 0 \,.
\ee
We can introduce the tangent space gamma matrices through
\be
\gamma^{\alpha} = e^{\alpha}{}_{M} \Gamma^M \,, 
\ee
where $e^{\alpha}{}_{M}$ is the vielbein.
The spinorial covariant derivative is then
\be
D_M = \pa_M + \half \w^{\a \b}_M \gamma_{\a \b} \,.
\ee

We now proceed to separate the spinor $\epsilon$ into various
components using projection operators. Firstly one can impose
the chirality condition
\be
\gamma_{11} \epsilon = - \epsilon \,,
\ee
where as usual $\gamma_{11} = -\gamma^{+-12r \cdots}$. Then we split
\be
\epsilon = \epsilon_+ + \epsilon_- \equiv \half (1 + i \gamma^{+-12}) \epsilon
+  \half (1 - i \gamma^{+-12}) \epsilon \,,
\ee
and furthermore split
\be
\epsilon_\pm = \epsilon_{\pm}^{(+)} + \epsilon_\pm^{(-)} \equiv - \half \gamma^- \gamma^+
\epsilon_\pm - \half \gamma^+ \gamma^- \epsilon_\pm  \,.
\ee
Using this decomposition it is straightforward to solve the Killing spinor
equations (\ref{eq:killing}) for the near horizon Schr\"odinger geometry
(\ref{eq:background3}). The result is
\be
\epsilon_\pm^{(+)} = 0 \,, \quad \epsilon_-^{(-)} = 0 \,, \quad \epsilon_ +^{(-)} =
\sqrt{r}\,  h(X_5)\, \eta_+^{(-)} \,,
\ee
where $\eta_+^{(-)}$ is a constant spinor consistent with the projections
above and $h(X_5)$ is an operator that projects onto Killing spinors on the internal
space $X_5$.

Introducing the wave therefore breaks any pre-existing symmetries down to
one quarter. In particular, if $X_5 = S^5$ the background (\ref{eq:background3})
is 1/4 BPS, while if $X_5 = T^{1,1}$ or any other Sasaki-Einstein space, the
background is 1/16 BPS.  The surviving symmetries are half of
the supertranslations. This suggests that
the supersymmetry algebra of this background may be the
algebra with eight supercharges found in \cite{Sakaguchi:2008rx}.
Other recent interesting discussions of a super Schr\"odinger symmetry
for AdS/CFT include \cite{Sakaguchi:2008ku, Nakayama:2008qm}.
A systematic approach to Schr\"odinger superalgebras was presented
in \cite{Duval:1993hs}.

Including a nonzero $B_2$ is likely to break the supersymmetry altogether,
as noted in \cite{Herzog:2008wg, Maldacena:2008wh}.

\section{A renormalisation group flow}
\label{sec:rgflow}

In this section we study a renormalisation group flow. An
interesting novelty compared with the well studied
Poincar\'e invariant renormalisation group flows in
AdS/CFT is that here the time and spatial parts of the metric
can scale differently as a function of the radial
distance (this fact has been used in
\cite{Gubser:2008wz} and \cite{Kachru:2008yh} to find flows
that interpolate between fixed points with different
values of the speed of light and $z$, respectively).
This allows more freedom in finding solutions.

The following relatively simple ansatz
is consistent:
\bea\label{eq:background6}
ds^2_{IIB} & = & R^2 \left( - 2 s(r) d\xp d\xm + t(r) d\xb^2 +
\frac{dr^2}{r^2} - f(X_5) m(r) (d\xp)^2 + p(r) ds^2_{X_5} \right) \,, \\
F_5 & = & 4 R^4 (1 + \star) \text{vol}(X_5) \,.
\eea
As before $-\nabla^2_{X_5} f = 12 f$. The Einstein equations
become a set of nonlinear equations for the four functions
$\{s,t,m,p\}$. There is a first order constraint
\be\label{eq:firstorder}
r^2 \left(\frac{1}{4} \frac{s'^2}{s^2} + \frac{1}{4}
\frac{t'^2}{t^2} + \frac{5}{2} \frac{p' s'}{p s} + \frac{5}{2} \frac{p' t'}{p t} + \frac{5}{2} \frac{p'^2}{p^2}
+ \frac{t' s'}{t s} \right) - \frac{10}{p} + \frac{4}{p^5} = 0 \,,
\ee
and three second order equations
\bea
m'' & = & \left(\frac{s'}{s} - \frac{t'}{t} - \frac{5}{2}
\frac{p'}{p} - \frac{1}{r} \right) m'
+ \left(- \frac{s'^2}{s^2} + \frac{12}{r^2 p} + \frac{8}{r^2 p^5} \right) m \,, \\
s'' & = & \left(- \frac{t'}{t} - \frac{5}{2} \frac{p'}{p} - \frac{1}{r} \right) s' + \frac{8 s}{r^2 p^5} \,, \\
t'' & = & \left(- \frac{s'}{s} - \frac{5}{2} \frac{p'}{p} -
\frac{1}{r} \right) t' + \frac{8 t}{r^2 p^5} \,.
\eea
From these equations and the first order equation
(\ref{eq:firstorder}) it is straightforward to obtain an equation
for $p''$. Note the structure of the equations: there are
three equations for $\{s,t,p\}$ and then a decoupled
equation for $m$. Therefore any solution will be a
(generally non-Poincar\'e invariant) renormalisation group
flow of the ${\mathcal{N}}=4$ theory that is then
deformed into a nonrelativistic background, without
backreaction on the initial metric functions.

Remarkably, it is possible to find an explicit analytic solution
to these equations. In fact, one can find the general solution
with a constant size $S^5$, that is $p=1$. The remaining radial
functions become
\bea
m(r) & = & \left(\frac{r^4-r_0^4}{r^4+r_0^4} \right)^{\sqrt{3}/2}
\frac{\sqrt{r^8 - r_0^8}}{r_0^8} \, E\left(1 - \frac{r_0^8}{r^8} \right) \,, \\
s(r) & = & \left(\frac{r^4-r_0^4}{r^4+r_0^4} \right)^{\sqrt{3}/2}
\frac{\sqrt{r^8 - r_0^8}}{r^2} \,, \\
t(r) & = & \left(\frac{r^4+r_0^4}{r^4-r_0^4} \right)^{\sqrt{3}/2}
\frac{\sqrt{r^8 - r_0^8}}{r^2} \,.
\eea
Here $E(k)$ is a complete elliptic integral of the second
kind\footnote{Recall that $E(k)=\int_0^{\pi/2} \sqrt{1- k^2
\sin^2\theta} d\theta$. The elliptic integral may also be expressed
in terms of hypergeometric functions or associated Legendre
functions. There seems to be a bug in (some versions at least of)
Maple in simplifying these elliptic integrals.}. Note that $E(0) =
\pi/2$. There is a second independent solution which has a
logarithmic divergence as $r \to r_0$, which we have not written
here.

The first fact we can note is that this metric asymptotes as $r
\to \infty$ to our original solution (\ref{eq:background3}).
Thus, we asymptotically develop a Schr\"odinger symmetry.
On the other hand, the solution is singular at $r=r_0$. It is
easy to check that the Riemann tensor squared diverges as $r \to
r_0$. However, it is a null singularity
(this is seen from the fact
that it takes an infinite boundary time for an infalling null observer
to reach the singularity)
and therefore likely to be
admissible as a solution in the full string theory \cite{Gubser:2000nd}.

If we had taken $m=0$, this flow would have a simple
interpretation. The functions $s$ and $t$ appearing in the metric
(\ref{eq:background6}) are dual to components of the energy
momentum tensor in the (3+1 dimensional) ${\mathcal{N}}=4$ theory.
There is a vacuum expectation value but no source for these modes.
Therefore we would be looking at the theory in a state in which
the expectation value for the pressure is anisotropic. This is not
especially natural in the absence of anisotropic sources.
However, once we have $m \neq 0$ there is a
different interpretation.

Viewed as a dual to a nonrelativistic theory, we identify the $\xm$
coordinate which is now no longer a spacetime
coordinate in the dual (2+1 dimensional) field theory.
According to \cite{Son:2008ye, Adams:2008wt},
the $r^2$ coefficient of $(d\xp)^2$ couples to the particle number
density of the dual theory while in our solution the subleading term
in $m(r)$ is a constant. Therefore it seems that this solution
corresponds to a finite energy density but zero particle density.
This statement is furthermore supported by the absence of a $(d\xm)^2$
term, which appeared in the finite particle density solutions of
\cite{Herzog:2008wg, Maldacena:2008wh, Adams:2008wt}.

Another consistent simplification of the equations above is to
set $s=t$. We have not been able to solve the resulting equations
analytically. This flow, with $m=0$,  would be a Poincar\'e
invariant state of the ${\mathcal{N}}=4$ theory and therefore
less specific to the nonrelativistic setup.

\section{Coulomb branch solutions}

The methods we have been using are easily adapted to produce
Coulomb branch solutions that asymptote to nonrelativistic
backgrounds. Our starting point will be the explicit supersymmetric
and asymptotically AdS backgrounds that were constructed in
\cite{Freedman:1999gk}. There are five such backgrounds, which
preserve an $SO(n) \times SO(6-n) $ subgroup of the $SO(6)$
symmetry of the ${\mathcal{N}}=4$ theory. The fact that these
backgrounds only involve the metric and five form field strength
make them especially amenable to non-relativistic generalisation.
Some solutions related to those we shall present shortly, but
with different asymptotics, were found in  \cite{Sfetsos:2005bi}.

\subsection{Asymptotically Schr\"odinger solution with $SO(2) \times SO(4)$ symmetry}

We will write down directly the Coulomb branch solution with the
nonrelativistic deformation.\footnote{In this section we set the AdS radius $R=1$ for simplicity.}
The metric and five form of our solution are
\bea\label{eq:c1}
ds^2  & = & \zeta r^2 \left( -2 d\xp d\xm + d\xb^2 + \frac{dr^2}{r^4 \lambda^6} - m(r,S^5) (d\xp)^2 \right)
\nonumber \\
& & \qquad \qquad + \, \zeta d\theta^2 + \frac{\cos^2\theta}{\zeta} d\Omega^2_{S^3} + \frac{\lambda^6 \sin^2\theta}{\zeta} d\Omega^2_{S^1}  \,, \nonumber \\
F_5 & = & \left[ r \left(L^2+4r^2+L^2\cos2\theta\right) dr - L^2 r^2 \sin 2\theta d\theta \right] \wedge d\xp \wedge d\xm \wedge dx \wedge dy \,, 
\eea
with the remaining components of $F_5$ obtained by Hodge dualising.
We introduced the functions
\be\label{eq:zeta}
\lambda^6=1+\frac{L^2}{r^2}\,, \qquad 
\zeta^2 = 1+ \frac{L^2}{r^2}\cos^2\theta\,.
\ee
Here $L$ gives a notion of the distance between the smeared $D3$ branes. At $L=0$
the solution reduces to a gravitational wave in pure AdS. Setting $m=0$ gives the solution in
\cite{Freedman:1999gk}. Introducing a nonzero $m$ will still solve the IIB equations
of motion (\ref{eq:IIBeqns}) provided that $m$ is harmonic on the spacetime with $m=0$,
i.e. $\nabla^2 m = 0$. Writing this out explicitly gives
\be\label{eq:meqn}
\frac{1}{r^3 \zeta} \frac{d}{dr} \left(\lambda^6 r^5 \frac{dm}{dr} \right)
+ \frac{1}{\zeta} \frac{1}{\cos^3\theta \sin\theta}
\frac{d}{d\theta} \left(\cos^3\theta \sin\theta \frac{dm}{d\theta} \right)
+ \frac{\zeta}{\cos^2\theta} \nabla^2_{S^3} m + \frac{\zeta}{\lambda^6 \sin^2\theta} \nabla^2_{S^1} m = 0 \,.
\ee
 
The differential equation (\ref{eq:meqn}) can be separated easily if we take $m$ to be
independent of the coordinates of the $S^3$ and $S^1$ in the spacetime (\ref{eq:c1}).
To find a regular solution on the $S^5$ we must therefore find spherical harmonics on $S^5$ that only depend on the coordinate $\theta$. This is the coordinate that arises if we parametrise $S^5 \subset \R^6$ by 
\be\label{eq:parametrise}
\vec x = \cos \theta \, \vec n_4 + \sin\theta \, \vec n_2 \,,
\ee
where $\vec n_4$ and $\vec n_2$ are unit vectors taking values in an orthogonal $\R^4$ and $\R^2$, respectively, in $\R^6$. Recall that spherical harmonics on $S^5$ with eigenvalue $\ell (\ell+4)$ are given by tracefree symmetric tensors of rank $\ell$ on $\R^6$. Requiring $SO(4)\times SO(2)$ invariance restricts us to tensors of the form
\be
c_{a_1 \cdots a_{2M}} x^{a_1} \cdots x^{a_{2M}} = \sum_{N=0}^M \alpha_{N} (x_1^2 + x_2^2 + x_3^2 + x_4^2)^N  (x_5^2 + x_6^2)^{M-N} \,.
\ee
The tracefree condition (after allowing for the symmetry in this tensor) then imposes $M$ relations between the $M+1$ different $\alpha_N$.
Thus we can find a spherical harmonic solution for any even $\ell=2M$. By our formula (\ref{eq:zell})
we thus obtain Coulomb branch solutions that asymptote to nonrelativistic backgrounds
for any positive integer $z \in \Z^+$. Let us write explicitly the solution in the case 
$\ell=2$. Using the parametrisation (\ref{eq:parametrise}) and furthermore solving the remaining radial
equation in (\ref{eq:meqn}) leads to the solution
\be
m_{(z=2)} = a \left(1 + \frac{3 r^2}{2 L^2} \right) \left(- \frac{2}{3} + \cos^2\theta \right) \,.
\ee
Here $a$ is a free parameter.
There may of course be other more complicated solutions to (\ref{eq:meqn}) in which the variables
do not separate. It is pleasing that the simplest regular solution turns out to have the properties we are interested in. The solution is a gravitational wave in a supersymmetric Coulomb branch geometry that asymptotes to our nonrelativistic dual spacetime (\ref{eq:background3}), with a particular choice for the $\ell = 2$ spherical harmonic in that solution.

\subsection{$SO(3) \times SO(3)$  and $SO(5)$ symmetric solutions}

The treatment of the other two possible symmetries is entirely analogous, and so we
shall be brief. We take the $SO(3)\times SO(3)$ case first. The solution is
\bea\label{eq:c2}
ds^2  & = & \zeta r^2 \lambda \left( -2 d\xp d\xm + d\xb^2 + \frac{dr^2}{r^4 \lambda^6} - m(r,S^5) (d\xp)^2 \right) \nonumber \\
 &  &\qquad \qquad + \, \frac{\zeta}{\lambda} d\theta^2 + \frac{\cos^2\theta}{\zeta \lambda} d\Omega^2_{S^2} + \frac{\lambda^3 \sin^2\theta}{\zeta} d\Omega^2_{S^2}  \,,  \\
F_5 & = & \left[ \frac{L^4 + 8 L^2 r^2 + 8 r^4 + L^2 (L^2+ 2 r^2) \cos2\theta}{2\sqrt{L^2+r^2}} dr \right. \nonumber \\
& & \qquad \qquad \left. - \, L^2 r \sqrt{L^2+r^2} \sin 2\theta d\theta \right] \wedge d\xp \wedge d\xm \wedge dx \wedge dy \,, 
\eea
with the remaining components of $F_5$ obtained by Hodge dualising.
While $\zeta$ is as before in (\ref{eq:zeta}), we now have
\be
\lambda^4=1+\frac{L^2}{r^2}\,.
\ee
As before we can find solutions with integer values of $z$. The solution that has
an asymptotic Schr\"odinger symmetry is
\be
m_{(z=2)} =  a \frac{L^4 + 8 L^2 r^2 + 8 r^4}{L^2 r \sqrt{L^2+r^2}} \left(-\frac{1}{2} + \cos^2\theta \right) \,,
\ee
where again $a$ is a free parameter.

For the case with $SO(5)$ symmetry, the solutions take the form
\bea\label{eq:c3}
ds^2  & = & \frac{\zeta r^2}{\lambda^3} \left( -2 d\xp d\xm + d\xb^2 + \frac{dr^2}{r^4 \lambda^6} - m(r,S^5) (d\xp)^2 \right)
+ \zeta \lambda^3 d\theta^2 + \frac{\lambda^3 \cos^2\theta}{\zeta} d\Omega^2_{S^5} \,, \nonumber \\
F_5 & = & \left[ \frac{r^2(3 L^4 + 12 L^2 r^2 + 8 r^4 + L^2 (3 L^2+ 2 r^2) \cos2\theta)}{2(L^2+r^2)^{3/2}} dr 
\right. \nonumber \\
& & \qquad \qquad \left. - \, \frac{L^2 r^3}{\sqrt{L^2+r^2}} \sin 2\theta d\theta \right] \wedge d\xp \wedge d\xm \wedge dx \wedge dy \,, 
\eea
with the remaining components of $F_5$ obtained by Hodge dualising. For this solution
we have
\be
\lambda^{12} = 1+\frac{L^2}{r^2}\,.
\ee
The solution with an asymptotic Schr\"odinger symmetry has
\be
m_{(z=2)} =  \left(a \left(\frac{5}{6} + \frac{r^2}{L^2} \right)
+ b \frac{\sqrt{r^2+L^2} \left(24 r^4 + 8 L^2 r^2 - L^4 \right)}{L^2 r^3}\right) \left(-\frac{5}{6} + \cos^2\theta \right) \,.
\ee
Unlike in the other cases, there are two parameters here, $a$ and $b$.
Both of these solutions have an asymptotic Schr\"odinger symmetry.
One of them has a divergence as $r \to 0$, but given that the background itself
is singular in this limit it is not clear that this is sufficient a pathology to
throw away the solution.
As before, there are also solutions for all other positive integer $z$.

There were two further metrics considered in \cite{Freedman:1999gk}. These
are related to the $SO(2) \times SO(4)$ and $SO(5)$ metrics presented above
by letting $L^2 \to - L^2$. The metric with $SO(5)$ symmetry and the sign of
$L^2$ inverted is believed to be nonphysical.

We found above that introducing our pp wave into $AdS_5 \times S^5$
broke all the supersymmetries except for half of the supertranslations.
The Coulomb branch solutions for ${\mathcal{N}}=4$
super Yang Mills theory break the
superconformal symmetries but preserve all the supertranslations
\cite{Freedman:1999gk}.
Therefore, it seems likely that the solutions we have found in this
section will preserve half of the supertranslation symmetries.

\section{Discussion}

In this paper we have presented several solutions to IIB
supergravity with features that seem likely to be of interest
as duals for nonrelativistic quantum critical theories in 2+1
dimensions. We initiated a study of basic features of these solutions
including stability, regularity and renormalisation flow generalisations.
In this discussion we note important open questions and comment on the dual
interpretation of these backgrounds.

Given that our backgrounds have (at least) a manifest 2+1 dimensional
Galilean symmetry, it seems clear that they are dual to nonrelativistic
theories defined on a 2+1 dimensional spacetime. In particular,
the mixing between the `AdS' and internal coordinates in the bulk
should not alter this fact. Furthermore, our $z=2$ solutions are
continuously connected to the embedding found in
\cite{Herzog:2008wg, Maldacena:2008wh, Adams:2008wt}.
It therefore seems likely that, as with those solutions, our
bakgrounds are dual to (differing) twisted DLCQs of
${\mathcal{N}}=4$ Super Yang-Mills theory. The dual
description of the solutions
with $z \neq 2$ is possibly less clear from this point of view.
However, we noted that they share with the $z=2$ cases the interpretation
as the near horizon limit of a branewave.

It is of interest for several reasons to generalise the solutions
we have found to a finite temperature and also a finite number
density, with a $(d\xm)^2$ term in the metric.
One reason is that most potential applications to
experimental systems will be at least at finite number density and
often at finite temperature. Another reason is that, as we have
mentioned several times above, in solutions with vanishing
temperature and number density the periodically identified circle
(\ref{eq:periodic}) is null. This is expected to invalidate the
supergravity approximation as wrapped strings can become light
\cite{Maldacena:2008wh}.

The mixing in our solution (\ref{eq:background3}) between the
noncompact and the $X_5$ directions makes it hard to find
generalisations with a $(dx^-)^2$ component in the metric. This is what
we would need to make the identified circle spacelike. In fact,
all of the solutions discussed in this paper
seem to be an application of the Garfinkle-Vachaspati
solution generating technique \cite{Garfinkle:1990jq,Garfinkle:1992zj}
(for a recent review of this method in a string theory context
see for instance \cite{Hubeny:2002nq}).
This method requires a null (hypersurface-orthogonal) Killing
vector field, which will not be present in finite temperature
solutions, or more generally if $\pa_{\xm}$ is spacelike. A
different approach will be required to find the finite temperature
solutions.

Restricting to the solutions we have found in this paper,
there are several immediate remaining questions. We did not analyze the
stability of the solutions with $z \neq 2$ and we did not consider the
stability of any of the backgrounds against other modes such as
gravitational perturbations. We also did not find the explicit form
of the unstable scalar mode.

It seems clear that our constructions can straightforwardly be generalised to
different spacetime dimensions. It would also be interesting to generalise
the renormalisation group flow solution of section \ref{sec:rgflow}
to include a $B_2$ field, and also to different values of $z$.

\section*{Acknowledgements}

It is a pleasure to acknowledge helpful discussions with Alejandra
Castro, Chris Herzog, John McGreevy, Mukund Rangamani
and Makoto Sakaguchi.
SAH was based at the KITP in Santa Barbara while much of this
work was done and furthermore acknowledges the stimulating
hospitality of the TIFR, Mumbai, while this work was being completed.
This research was  supported in part by the National Science Foundation
under Grant No. PHY05-51164. The work of KY was also supported in part by
JSPS Postdoctoral Fellowships for Research Abroad.

\end{document}